\documentclass{article}
\usepackage[utf8]{inputenc}
\usepackage{authblk}
\usepackage{graphicx}
\usepackage{color}
\usepackage{amsmath}
\usepackage{amssymb}
\usepackage{epstopdf}
\usepackage[english]{babel}
\usepackage{caption}
\usepackage{subcaption}

\title{Active Liquid-Liquid Phase-Separation in a Confining Environment.}
\author[1,2,4]{Chen Lin}
\author[3]{Robijn Bruinsma}
\affil[1]{Center for Computational Biology, Flatiron Institute, New York}
\affil[2]{Center for Computational Mathematics, Flatiron Institute, New York}

\affil[3]{Department of Physics and Astronomy, University of California, Los Angeles}
\affil[4]{Department of Chemistry and Biochemistry, University of California, Los Angeles}

\date{\today}

\begin{document}

\maketitle

\begin{abstract}
    
\end{abstract}
Active liquid-liquid phase separation (LLPS) in a confining environment is believed to play an important role in cell biology. Recently, it was shown that when active noise at the microscopic level is included in the classical theory of nucleation and growth then this does not cause the breakdown of detailed balance at the \textit{macroscopic} level provided that the droplet radius is the only collective coordinate. Here, we present a simple model for active LLPS in a confining environment, with the droplet location in a confining potential as a second collective coordinate, and find that detailed balance \textit{is} broken at the macroscopic level in an unusual fashion, using the Fluctuation-Dissipation Theorem as a diagnostic.

\section{Introduction}

Liquid-liquid phase separation (LLPS) is, in its simplest form, a phase transition where a homogeneous, binary liquid mixture segregates into two parts, each enriched in one of the two constituent components \cite{falahati2019thermodynamically}. LLPS may be initiated by a rapid change in the thermodynamic parameters that transforms a thermodynamically stable, homogeneous state into an unstable supersaturated state. This initiates a process of nucleation of minority phase droplets, their subsequent growth and their merger \cite{bray1994theory}, terminating, again in the simplest case, in a final state of phase coexistence composed of two thermodynamically stable, macroscopic components separated by a planar phase boundary. 

LLPS currently is receiving much interest in the context of cell biology \cite{shin2017liquid,hyman2014liquid}. For example, cells may respond to changes in their internal physico-chemical parameters by concentrating certain macro-molecules into droplet-like structures, known as stress granules or also "organelles" (unlike the typical organelles of cells, these are not surrounded by a lipid membrane).\footnote{Because of their lack of optical contrast, membrane-less organelles appear to have been overlooked in the past although they were apparently observed as early as 1957 in electron microscopy images \cite{andre1957ultrastructure}.}  Another example are the P granules of \textit{C. Elegans} that appear at the nuclear envelope of germline cells. These organelles contain proteins associated with RNA molecules, producing insoluble aggregates \cite{boeynaems2018protein}.  

Cell-biological LLPS has important features that sets it apart from conventional LLPS. First, the formation of P granules is known to be a highly dynamical process involving ATP-consuming enzymes \cite{chen2020dynamics}. The classical theory of nucleation and growth \cite{lifshitz1961kinetics} is based on the fundamental principle of detailed balance (PDB), itself a consequence of the time-reversal symmetry of the underlying microscopic interactions. The PDB provides important relations between the on and off-rates of the different kinetic processes of a statistical mechanical system assuring that, in the long-time limit, the system evolves to a state of thermodynamic equilibrium. \textit{In general} the PDB does not apply to systems that actively consume free energy, such as the production of P granules  \cite{cates2012diffusive}.

A second feature that sets apart the biological version of LLPS is that it does not terminate in a state of macroscopic phase coexistence. Instead, the size of the organelles does not exceed the range of a few microns. One possible reason, which we will assume to apply, is the increasing energy cost of deformation of the cytoskeleton surrounding a swelling organelle \cite{liu2023liquid}. This feature has been reproduced under \textit{in-vitro} conditions when passive LLPS is instigated in a binary mixture of two simple fluids permeating a polymer gel such that the polymers of the gel are insoluble in the minority component of the mixture \cite{fernandez2021putting}. The deformation energy of the cytoskeleton is expected to be strongly non-linear with respect to the organelle size because of strain hardening of the cytoskeleton \cite{xu2000strain}. The confinement effect allows for a non-equilibrium steady state (NESS) in which the nucleation and evaporation of droplets balance each other.  

The motivation of this paper is a surprising aspect of the theory of active nucleation. It has been shown that when active noise at the molecular-level is introduced in the classical theory of nucleation and growth then this does \textit{not} lead to observable violations of the PDB at the macroscopic level \cite{cates2023classical} \textit{provided the description of the macroscopic dynamics includes only the droplet radius as a collective coordinate}. Active nucleation in a confining environment that prevents the growth of droplet-like organelles may be an interesting laboratory to further investigate this fundamental issue. On the one hand due to the lack of translational symmetry, one minimally needs to include the displacement of the droplet as an additional collective variable. On the other hand, if the confinement sufficiently reduces the droplet size then the time-reversed process, i.e., droplet evaporation, will start to become noticeable. Under a non-equilibrium steady-state (NESS) such that the number of nucleation events is balanced by the number of evaporation events, the observation of many such events could allow experiments to arrive at quantitative conclusions about the applicability of the PDB.

We propose here a very simplified model for active droplet nucleation and evaporation in a confining environment. The model includes \textit{two} collective degrees of freedom, namely the droplet radius and the droplet displacement. The aim of the model is not to provide even an approximately realistic description of the active nucleation of, say, P granules but rather to explore a tractable minimal model that adds to the primary radius degree of freedom of the classical theory one other collective degree of freedom. In the model, droplets are initiated by an active nucleation source located at the center of a spherically symmetric potential trap. After the droplet is initiated at the origin and starts to grow, it is displaced from the origin by Brownian motion. Displacement from the origin increases the chemical potential of the molecules inside the droplet, thus making a droplet moving away from the origin more prone to evaporation. Since no droplets are allowed to nucleate away from the origin, time-reversal symmetry is explicitly violated. The model is defined in more detail in Section II where we also discuss the ``mean-field" droplet dynamics. In Section III we inspect numerically computed time series for the droplet radius and displacement for different levels of confinement of the droplet and the rate of active droplet initiation. We then test the Fluctuation-Dissipation Theorem (FDT) that should hold for passive but not for active systems \cite{das2020introduction}. In Section IV we interpret our findings in the context of stochastic resetting processes.

\section{Model}

The rate equations for the droplet radius and displacement of the proposed model are: 
 \begin{align}
    & \frac{dR}{dt}=\frac{vD}{R}\Bigl(c_{\infty} - c_s(R,r)\Bigr)+\eta(t)\\&\gamma\frac{d{r}}{dt} = -\frac{\partial \Delta U(R,r)}{\partial  r}+ { \zeta}(t)
\end{align}
where $R(t)$ is the radius of a spherical droplet of minority phase molecules in an otherwise homogeneous supersaturated binary mixture and $ r(t)$ is the displacement of the droplet from the origin. To maintain simplicity, the displacement is treated here as a one-dimensional coordinate. Equation 1 is Fick's First Law under conditions of spherical symmetry, with $D$ here the diffusion coefficient of the minority phase molecules and $v$ the inverse of the density of a minority-phase liquid. The diffusion current to (or from) the droplet is proportional to the difference between the concentration of the minority phase $c_{\infty}$ far from the droplet and the concentration $c_s(R,r)$ at the surface of the droplet. For $c_s(R,r)$, we assume the Ostwald-Freundlich relation $c_s(R,r)=c_{eq}\exp\big(\Delta P(R,r) v/k_bT\big)$ with $\Delta P$ the excess pressure inside the droplet as compared to the surrounding fluid and with $c_{eq}$ the saturation concentration, i.e., the minority phase solution concentration when the solution and the minority phase liquid state are in phase coexistence. The difference $S=(c_{\infty}-c_{eq})/c_{\infty}$ is a measure of the degree of supersaturation. In classical nucleation theory, $\Delta P$ is equated to the Laplace pressure $2\sigma/R$ inside the drop with $\sigma$ the interfacial energy between the droplet interior and the surrounding medium\footnote{If the exponential of the Ostwald-Freundlich is expanded to lowest order in its argument then Eq.\ref{eq1} reduces to the expression for droplet growth rate that was used by Lifshitz and Slyozov in their classical study of droplet nucleation and growth \cite{lifshitz1961kinetics}.} but for the present case, we include in $\Delta P$ also the pressure generated by an external, spherically symmetric parabolic confining potential acting on the minority-phase molecules in the droplet. This leads to a total potential energy
$\Delta U(R,r)$
\begin{equation}
\Delta U(R,r)=4\pi R^2 \sigma+\left(\frac{4\pi}{5v}\right) \alpha  [R^5+(5/3) R^3 r^2]   
\end{equation}
with $\alpha$ a coefficient with dimensions of energy over length squared that measures the strength of the potential\footnote{$\alpha$ would be a measure of the degree of strain-hardening of the cytoskeleton.}. From the relation $\Delta P = \frac{d\Delta U}{dV}$, with $V=\frac{4\pi R^3}{3}$ the droplet volume, one finds for the excess pressure 

\begin{equation}
    \Delta P(R,r) v /k_b T= \frac{2\sigma v}{R}+(\alpha/k_bT) \left[R^2 + r^2\right].  
\end{equation}
The last term $\eta(t)$ of Eq.1 is a noise source that will be specified below. Equation 2 is a standard Langevin equation for the motion of the droplet in the potential $\Delta U(R,r)$ with $\gamma$ the friction coefficient\footnote{The friction coefficient of a small sphere in a fluid depends on $R$. This is not included in the rate equations as this complicates the numerical integration of the rate equations when the radius is very small.} and with ${\zeta}(t)$ a noise source that will be specified below. 

It will be useful to connect the rate equations with the droplet Gibbs free energy $\Delta G(R,r)$, given by
\begin{align}
 \Delta G(R,r)=-\bigg(\frac{4\pi R^3}{3v}\bigg)k_b T\ln\left(\frac{c_{\infty}}{c_{eq}}\right)+\Delta U(R,r)
\end{align}
 The first term is the reduction of the entropic free energy of the system when a (nearly) single-phase droplet forms in a supersaturated mixture. It is easy to show that in the limit of low supersaturation and $\Delta P(R,r) v /k_b T$ is small compared to one then one can approximate the rate equations as a pair of coupled Langevin-type rate equations
\begin{align}
    & \frac{dR}{dt}\simeq-\bigg(\frac{v^2Dc_{\infty}}{4\pi R^3 k_bT}\bigg)\frac{\partial \Delta G(R,r)}{\partial  R}+\eta(t)\label{eq1}\\&\gamma\frac{d{r}}{dt} = -\frac{\partial \Delta G(R,r)}{\partial  r}+ { \zeta}(t)\label{eq2}
\end{align}
The two rate equations have noise sources associated with the radius and displacement. As discussed in Section I, the nucleation of granules is an active process. To keep the model as simple as possible, we will assume that the noise sources in the two rate equations do not violate the PDB. In that case the associated noise-noise correlation functions can be determined heuristically by formally assuming that $\Delta G(R,r)$ is harmonic in both degrees of freedom, so the two rate equations are linear and can be integrated. Imposing the condition that the steady-state probability distribution is the Boltzmann distribution leads to the conditions $<\zeta(t)\zeta(t')>=2\gamma k_bT\delta(t-t')$ and  $<\eta(t)\eta(t')>|_R=\Delta_{\eta}(R)\delta(t-t')$ with $\Delta_{\eta}(R) = \left(\frac{v^2 D c_{\infty}}{2\pi R^3}\right)$. Note how the noise intensity for the radius diverges in the limit that the droplet radius goes to zero. This must be the general answer assuming that the thermal noise in a solution does not depend on the form of the Gibbs free energy\footnote{For a legitimate discussion, see ref.\cite{cates2023classical}}. Finally, we must specify the active droplet nucleation and evaporation process. We will assume that there is an active nucleation center located at the origin of the potential. If there is no droplet present at time t=0 then a droplet will be introduced between times $t$ and $t+dt$ with probability $p(t)dt$ where $p(t)=\frac{1}{\tau} \exp (-t/\tau)$ with $\tau$ the mean waiting time for the appearance of a fresh droplet. 

 \section{Landau Theory}
\begin{figure}
    \centering
    \includegraphics[scale=0.4]
    {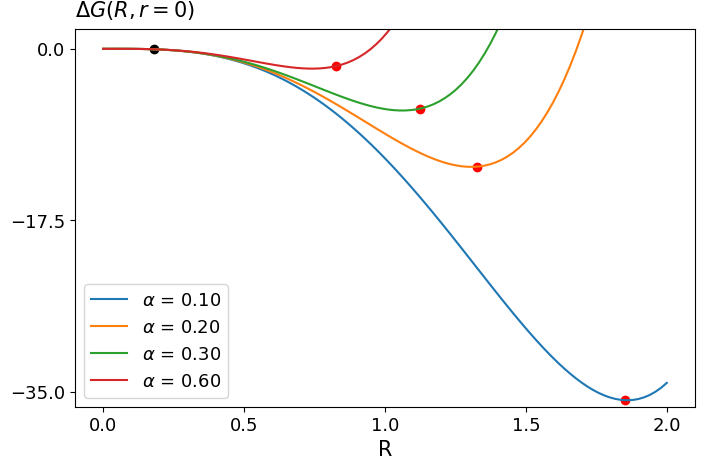}\caption{Plot of the droplet free energy $\Delta G(R,0)$ in units of $k_BT$ as a function of the droplet radius $R$ in units of the molecular length scale $v^{1/3}$ for different values of the parameter $\alpha$ = 0.1, 0.2, 0.3, and 0.6 expressed in units of $k_BT/v^{2/3}$. The red and black dots on the curves show, respectively, the minimum $R^+$, which corresponds to a stable droplet, and the maximum $R^-$, which corresponds to the critical nucleus radius. The surface energy is $\sigma$ = 0.1 in units of $k_BT/v^{2/3}$, $c_{\infty} v = 0.1$ and  $c_{eq}/c_{\infty} = 0.7$.}\label{delta_G}
\end{figure}

 If one drops the two thermal noise sources from the stochastic rate equations \ref{eq1} and \ref{eq2} then the dynamics is determined by gradients of the Gibbs free energy. Figure \ref{delta_G} is a plot of $\Delta G(R,0)$ for the case of a drop fixed at the origin and for different values of the confinement parameter $\alpha$.
The Gibbs free energy always has a minimum at the empty state $R=0$. For low values of $\alpha$ in units of $k_BT/v^{2/3}$ there is, in addition, a maximum at $R^-\simeq \frac{2\sigma v}{k_b T\ln{\left(\frac{c_{\infty}}{c_{eq}}\right)}}$ followed by a second minimum at $R^+\simeq \left[\frac{k_b T}{\alpha}\ln{\left(\frac{c_{\infty}}{c_{eq}}\right)}\right]^{1/2}$.
The maximum at $R^-$ is the critical radius of classical nucleation theory. Because the surface energy of organelles is believed to be low, we will assume that the interfacial energy $\sigma$ is small compared to one in the natural units $k_BT/v^{2/3}$  (in Fig.1, $\sigma = 0.1$). As Fig.1 shows, the activation energy maximum at the critical radius is then very shallow. Note that the critical radius $R^-$ is, in that case, less than the molecular length scale $v^{1/3}$ except for very low values of the supersaturation. As shown in Fig.1, for low values of $\alpha$, the Gibbs free energy has a deep minimum at $R^+$, which corresponds to a stable, static droplet. As $\alpha$ increases, the stable droplet radius decreases and starts to approach $R^-$. The activation energy barrier for droplets starting at $R^+$ to cross the transition state at $R^-$ becomes progressively smaller. At a critical value of $\alpha$, a weakly first-order transition takes place when $\Delta G(R^+,0)=0$. For the case shown in Fig.1, this is near $\alpha\simeq 0.6$. This transition is closely followed by a spinodal point where $R^+$ and $R^-$ merge. Beyond this point, the $R=0$ minimum is the only extremum of the Gibbs free energy. 

\begin{figure}
    \centering
    \includegraphics[scale=0.6]
    {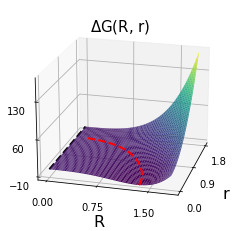}\caption{Plot of the droplet free energy $\Delta G(R,r)$ in units of $k_BT$ as a function of the droplet radius $R$ and droplet displacement $r$ for $\alpha=0.2$ and $\sigma=0.1$ in dimensionless units (see caption Fig.1). There is an absolute minimum at $R=R^+$ and $r=0$. The red dashed line that ends at the minimum $R=R^+,r=0$ is the locus of points along which $\frac{\partial\Delta G(R,r)}{\partial R}=0$. The black dashed line indicates points on the free energy surface with radius $R_0=0.1$. During simulations, droplets with radius less than $R_0$ were removed. Contours of equal $\Delta G(R,r)$ are shown as shaded black lines. Note the saddle-point.} \label{delta_G_3d}
\end{figure}

Next, allow for general $R$ and $r$. Figure \ref{delta_G_3d} shows $\Delta G(R,r)$ in the extended $(R,r)$ plane. 
The absolute minimum of $\Delta G(R,r)$ is at $R=R^+$ and $r=0$. According to Eqs.6 and 7,
a ``phase point" $(R(t), r(t))$ moves in the $(R,r)$ plane along flow lines such that tangents to the flow lines make a fixed angle with the gradients of $\Delta G(R,r)$. For the special case that $\frac{\gamma v^2 D c_{\infty}}{4\pi R^3 }=1$, this angle is zero in which case the flow lines are directed along the gradients of $\Delta G(R,r)$. The dynamics of $(R(t), r(t))$ is then the same as that of a mass point sliding down the potential hill $\Delta G(R,r)$. 

The derivative $\frac{\partial\Delta G(R,r)}{\partial R}$ vanishes along the red dashed line such that $\frac{\partial\Delta G(R,r)}{\partial R}$ is negative inside this line and positive outside. Since $\frac{\partial\Delta G(R,r)}{\partial r}$ is always positive, it follows that phase points that at $t=0$ are somewhere inside the red dashed line will stay inside the line and always move towards larger $R$ and smaller $r$. These flow lines must end up at the absolute minimum at $R^+$ and $r=0$, which is a stable fixed point of the dynamical system Eqs.\ref{eq1} and \ref{eq2}. For small $\sigma$ the red dashed line is circular and given by
\begin{equation}
R^+(r)\simeq \left[\frac{k_bT}{\alpha}\ln{\left(\frac{c_{\infty}}{c_{eq}}\right)}-r^2\right]^{1/2}
\label{R+}
\end{equation}
Flow lines emerging from initial coordinates outside the red dashed line still can flow towards the fixed point if $r(t=0)$ is small while they flow towards $R=0$ for larger values of $r(t=0)$. The separatrix between these two regimes (not shown) runs from the maximum of $\Delta G(R,r)$ in Fig.2 to the saddle point of the equal height contour lines shown in Fig.2.

The line $R^-(r)$ of critical radii is, for $\sigma=0.1$, significantly smaller than the molecular length $v^{1/3}$, as already noted. For the numerical simulations in the next section, $R^-(r)$ was, for that reason, replaced by the black dashed line in Fig.2 at $R_0=0.1$ that acted as the microscopic cutoff. Phase points with $R$ less than $R_0$ were removed from the system. Active droplet production at $r=0$ was assumed to generate droplets with radius $R_0$. The red dashed line crosses the cut-off line near $r^*\simeq\left[\frac{k_bT}{\alpha}\ln{\left(\frac{c_{\infty}}{c_{eq}}\right)}\right]^{1/2}$. For the flow lines, the black dashed line is a line of sources for $r$ less than $r^*$ and a line of sinks for $r$ greater than $r^*$. 

For fixed radius $R=R^+$ the displacement degree of freedom is a harmonic spring with spring constant $k(R^+)\simeq
\left(\frac{8\pi}{3v}\right)\alpha  {R^+}^3$, friction coefficient $\gamma$ and relaxation rate $k(R^+)/\gamma$. The spring constant decreases with $\alpha$ as $1/\alpha^{1/2}$, as does the relaxation rate, while the mean square $<r^2>=\frac{k_bT}{k(R^+)}$ for displacement fluctuations increases as $\alpha^{1/2}$. Next, for fixed displacement $r=0$, the radius degree of freedom close to the fixed point can also be approximated as a harmonic spring. The spring constant $K=\frac{8\pi}{\alpha^{1/2}v}\left(k_b T \ln\frac{c_{\infty}}{c_{eq}}\right)^{3/2}$ is again  proportional to $1/\alpha^{1/2}$. The effective friction coefficient is $\Gamma=\bigg(\frac{4\pi {R^+}^3 k_b T}{v^2Dc_{\infty}}\bigg)$ so the relaxation rate $K/\Gamma$ now is proportional to $\alpha$. 

Active noise was assumed to play no role in this section. Below, we will refer to this assumption as ``Landau theory". Landau theory is expected to be a good approximation as long as thermal fluctuations do not cause trajectories $(R(t),r(t))$ to cross the red dashed line and then end up at the evaporation line. This should be only rarely the case as long as the mean square $<r^2>$ of the displacement fluctuations around the fixed point is small compared to ${r^*}^2$. Since the first increases with $\alpha$, while the second decreases with $\alpha$, Landau theory should break down with increasing $\alpha$ for an $\alpha$ value of the order of one.

\section{Testing the Fluctuation-Dissipation Theorem.}

In this section, we numerically integrate the stochastic rate equations and test the FDT. Prior to numerical integration, the full rate equations were first expressed in dimensionless form as
\begin{align}
    & \frac{dR}{dt}=\frac{c_{\infty}}{R}\biggl(1 - (c_{eq}/c_{\infty})\thinspace e^{\thinspace \Delta P(R,r)}\biggr)+\eta(t)\label{eq11}\\&\gamma\frac{d{ r}}{dt} = -\frac{\partial \Delta U(R,r)}{\partial r}+ {\zeta}(t)\label{eq22}
\end{align}
The dimensionless time variable is the original time variable divided by $(v^{2/3}/D)$. The dimensionless distance variable is the original distance variable divided by $v^{1/3}$, and the new pressure is the old pressure divided by $k_BT/v$. Similarly, the constants $\alpha$ and $\sigma$ were made dimensionless by dividing by $k_B T/v^{2/3}$, the friction coefficient $\gamma$ by dividing by $k_B T/D$, and concentrations by dividing by $1/v$. In these units $\Delta_{\zeta} = 2\gamma$, $\Delta_{\eta}(R) = \left(\frac{c_{\infty}}{2\pi R^3}\right)$, $\Delta U(R,r)=4\pi R^2 \sigma+\left(\frac{4\pi}{5}\right) \alpha  [R^5+(5/3)R^3 r^2]$ 
and $\Delta P(R,r) = \alpha \left[R^2 +  r^2\right] +\frac{2\sigma}{R}$.

\begin{figure}
     \centering
     \begin{subfigure}[b]{0.45\textwidth}
         \centering         \includegraphics[width=\textwidth]{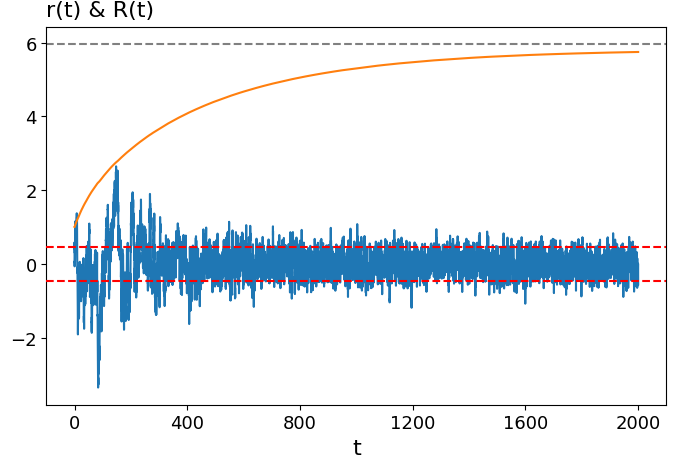}
         \caption{$\alpha = 0.01$, $\tau=1.0$}
         \label{fig:trj_00.1}
     \end{subfigure}
     \hfill
     \begin{subfigure}[b]{0.45\textwidth}
         \centering         \includegraphics[width=\textwidth]{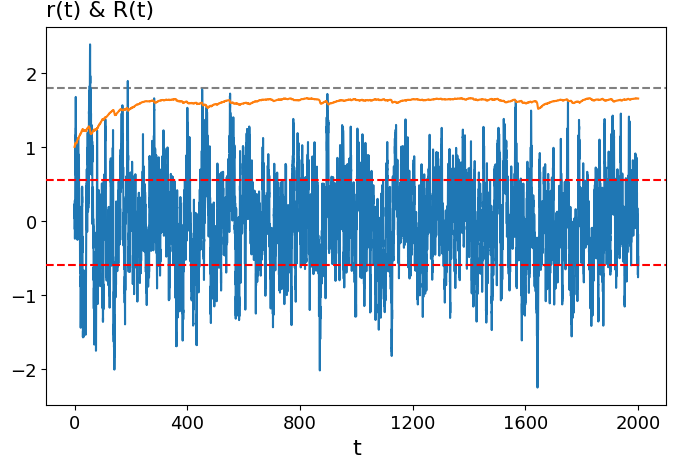}
         \caption{$\alpha = 0.1$, $\tau=1.0$}
         \label{fig:trj_0.1}
     \end{subfigure}
     \hfill
     \begin{subfigure}[b]{0.45\textwidth}
         \centering
         \includegraphics[width=\textwidth]{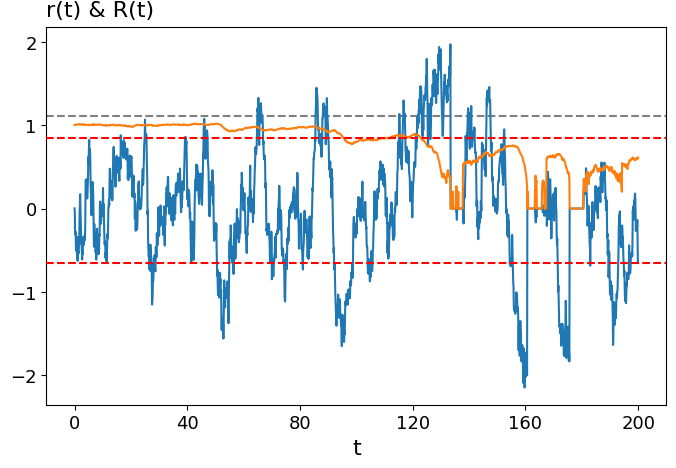}
         \caption{$\alpha = 0.2$, $\tau=1.0$}
         \label{fig:trj_0.2}
     \end{subfigure}
     \hfill
     \begin{subfigure}[b]{0.45\textwidth}
         \centering
         \includegraphics[width=\textwidth]{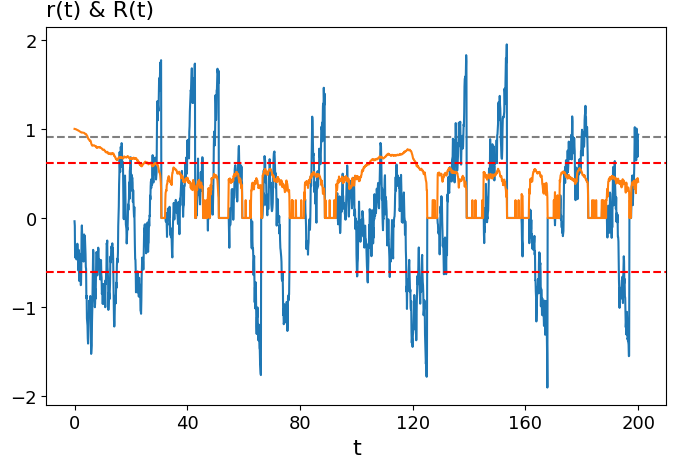}
         \caption{$\alpha = 0.3$, $\tau=1.0$}
         \label{fig:trj_0.3}
     \end{subfigure}
     \hfill
     \begin{subfigure}[b]{0.45\textwidth}
              \centering
    \includegraphics[width=\textwidth]
    {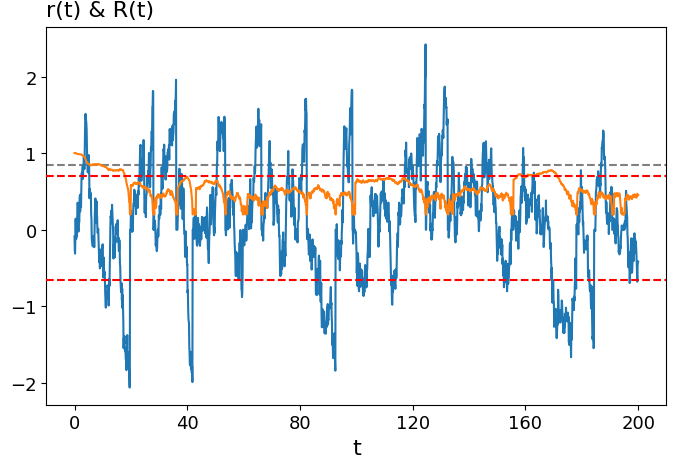} 
    \caption{$\alpha = 0.3$, $\tau=0$}
    \label{trj_ap0.3_tau0}
     \end{subfigure}
        \caption{The first four figures show time series of the radius and displacement for increasing $\alpha$ while the mean waiting time for droplet generation was $\tau=1.0$. The orange line is the droplet radius $R(t)$ and the blue line is the droplet displacement  $r(t)$. Radius and displacement are both expressed in units of the radius $v^{\frac{1}{3}}$. The black dashed line shows Eq.\ref{R+} for the stable radius $R^+$ while the red dashed lines are plus and minus the RMS displacement of the data. The last figure is again for $\alpha=0.3$ but now with the waiting time reduced to zero. For these, and all later plots, $\gamma=10$, $c_{\infty}=1$ and $c_{eq}=0.7$.}
        \label{fig:four regimes of dynamics}    
\end{figure}

Examples of numerically computed time series of the radius and displacement are shown in Fig.\ref{fig:four regimes of dynamics} for different values of the confinement parameter $\alpha$. 
The starting displacement was $r(0)=0$ while the starting radius was $R(0)=1$. Figure \ref{fig:trj_00.1} shows the case of $\alpha=0.01$. Starting at $t=0$, the droplet radius grows monotonically and then saturates near the black dotted line, which is the radius of a droplet given by Eq.\ref{R+} but with $r^2$  replaced by the mean square displacement of a harmonic spring with potential energy $ u_{R^+}(r) = \alpha\left(\frac{4\pi {R^+}^3}{3}\right) r^2$.  The blue line shows the thermal fluctuations of the displacement variable. The two red dashed lines are plus and minus the RMS displacement of $r(t)$. The initial fluctuations of $r(t)$ are significantly larger than at later times because the droplet radius initially is smaller than $R^+$. Recall that the displacement spring constant is proportional to ${R^+}^3$. There are no evaporation events. Thermal fluctuations of the radius are negligible.

Next, Fig.\ref{fig:trj_0.1} shows what happens if $\alpha$ is increased by an order of magnitude to $\alpha=0.1$. The average size of the droplet radius has decreased by a factor of about 0.3, which is close to the factor of $1/10^{1/2}$ predicted by Landau theory. The relaxation time of the droplet radius has decreased by a factor of about 10 (from about 500 to about 50), which is also consistent with Landau theory. Radius fluctuations are now noticeable. The RMS of the displacement fluctuations is now close to that of Landau theory because there is no longer an initial interval with droplet radius small compared to $R^+$ as in Fig.3a. There still are no evaporation events. 

Next, $\alpha$ was increased to $0.2$ (see Fig.\ref{fig:trj_0.2}). For the first 90 or so time steps, the radius is relatively close to Eq.\ref{R+}n (black dashed line). Around $t=90$ the droplet displacement experiences an unusually large thermal fluctuation, first to positive $r$ and then to negative $r$. This causes a reduction of the radius, consistent with the discussion in Section II. Then, around $t=125$, there is an even larger displacement event during which the droplet radius drops sharply, leading to a first evaporation event. Next, there are a few attempts at droplet nucleation but they fail. This is followed by a successful nucleation event around $t=135$. After another twenty or so time steps though, another large displacement deviation event causes evaporation of the new droplet. Over longer time intervals these two scenarios appear to alternate, suggesting a form of dynamical phase coexistence between a state with a large droplet radius and small displacement fluctuations, consistent with Landau theory, and a state in which droplets constantly evaporate and then briefly re-appear with large displacement fluctuations. When $\alpha$ is increased to $0.3$ (see Fig.3d), the system appears permanently trapped in the second state of Fig.3c with frequent evaporation/nucleation events and large displacement fluctuations. 

Finally, Fig.\ref{trj_ap0.3_tau0} shows the effect of reducing the waiting time to zero for $\alpha=0.3$. After a droplet evaporates, it re-appears nearly immediately. Notice the correlation between such events and displacement spikes of the droplet.

\subsection{FDT}
\begin{figure}
     \centering
     \begin{subfigure}[b]{0.45\textwidth}
         \centering
         \includegraphics[width=\textwidth]{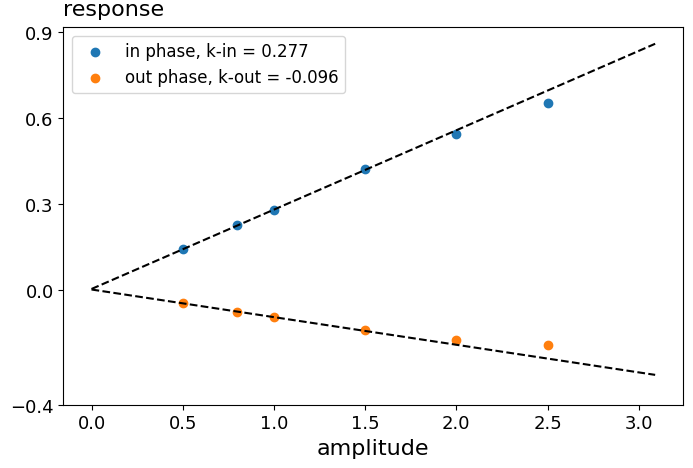}
         \caption{$\alpha = 0.1$}
         \label{fig:ap0.1_o0.1_curve}
     \end{subfigure}
     \begin{subfigure}[b]{0.45\textwidth}
         \centering
         \includegraphics[width=\textwidth]{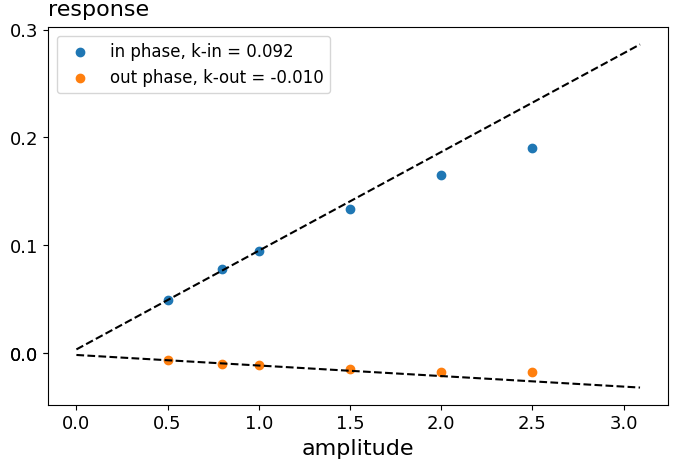}
         \caption{$\alpha = 0.3$}
         \label{fig:ap0.3_o0.1_curve}
     \end{subfigure}
     \hfill
     
        \caption{In-phase (blue dots) and out-of-phase (orange dots) response amplitudes of particle displacement, measured against the driving force amplitude at a frequency of $\omega = 0.1$. In both instants, $\tau=3.0$. For the left figure, $\alpha=0.1$ when nucleation and evaporation play no role (see Fig.3a). The fitted slopes of the in-phase and out-of-phase response are provided in the inset.  For the right figure, $\alpha=0.3$ there are many nucleation and evaporation events (see Fig.3d). Linear response breaks down for larger drive amplitudes significantly. The fitted slopes for low drive amplitudes are shown. Note the decrease of the linear response susceptibility as compared with $\alpha=0.1$.}
        \label{fig:curvature}
\end{figure}

For systems in thermodynamic equilibrium, the FDT relates the imaginary part of the linear response susceptibility $\chi{''}(\omega)$ in the frequency domain to the Fourier Transform $S(\omega)$ of the auto-correlation function (ACF), which is the fluctuation power spectrum:
\begin{equation}
 S(\omega)=-\frac{2}{\omega}\chi{''}(\omega).  
\end{equation}
We tested the FDT by including in the rate equations a mechanical force conjugate to the displacement $r(t)$. Figure \ref{fig:ap0.1_o0.1_curve} and \ref{fig:ap0.3_o0.1_curve} show the in-phase and out-of-phase response amplitudes of the displacement variable $r(t)$ as a function of the amplitude of a periodic driving force. The drive frequency $\omega$ was kept fixed at $\omega = 0.1$. 
For $\alpha=0.1$ the response is close to linear except for the highest amplitudes but for $\alpha=0.3$, non-linearity starts to appear at intermediate amplitudes. At those force amplitudes, a cubic harmonic at $3\omega$, appears in the power spectrum (not shown). This is reasonable because integrating out the radius fluctuations of $R(t)$ around $R^+$ generates a cubic non-linearity in the rate equation for $r(t)$. The FDT is restricted to linear response so the force amplitude needs to remain small enough so these signatures of non-linearity can be safely ignored. Note the decrease of the displacement response susceptibility when $\alpha$ is increased from $0.1$ to $0.3$. This conflicts with Landau theory since the spring constant decreases as $\alpha ^{-1/2}$ so the susceptibility should actually \textit{increase}. More generally, the susceptibility should grow large in the vicinity of the weakly first-order transition predicted by Landau theory

\begin{figure}
     \centering
     \begin{subfigure}[a]{0.45\textwidth}
         \centering
         \includegraphics[width=\textwidth]{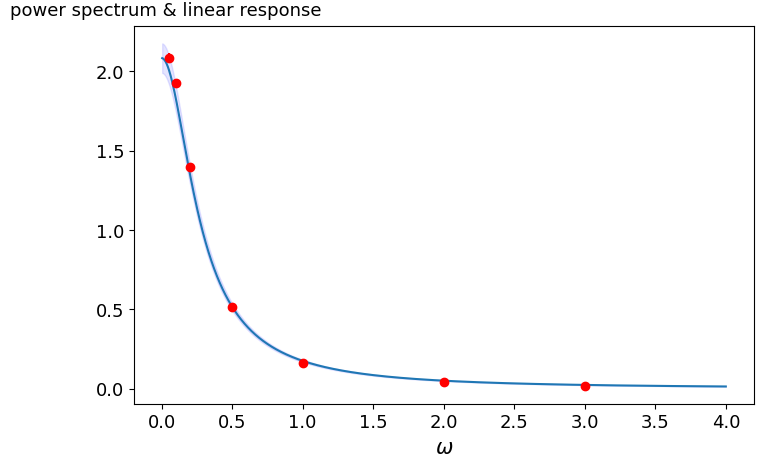}
         \caption{$\alpha = 0.1$, $\tau = 3.0$}
         \label{fig:0.1_3.0 full fdt}
     \end{subfigure}
    
     \begin{subfigure}[b]{0.45\textwidth}
         \centering
         \includegraphics[width=\textwidth]{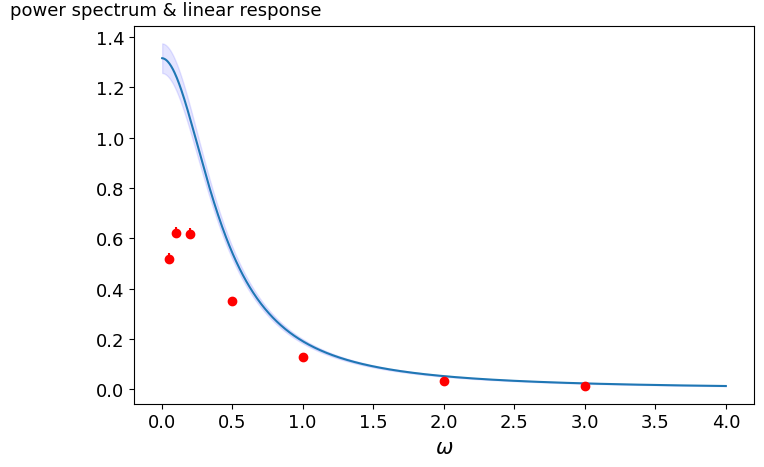}
         \caption{$\alpha = 0.2$, $\tau = 3.0$}
         \label{fig:0.2_3.0 full fdt}
     \end{subfigure}

     \begin{subfigure}[b]{0.45\textwidth}
         \centering
         \includegraphics[width=\textwidth]{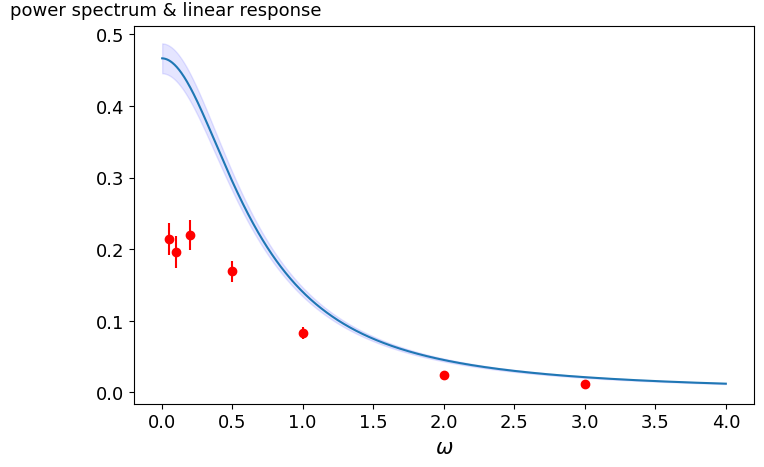}
         \caption{$\alpha = 0.3$, $\tau = 3.0$}
         \label{fig:0.3_3.0 full fdt}
     \end{subfigure}
    
     \hfill
        \caption{Testing the Fluctuation Dissipation Theorem (FDT) for the displacement variable. The confinement parameter $\alpha$ is increased from $0.1$ to $0.3$ while the initiation waiting time $\tau$ was kept fixed at $3.0$. The horizontal axis is the frequency $\omega$ and the vertical axis is the power spectrum $S(\omega)$ (blue line, the blue shading indicates error bars.) or $-\frac{2}{\omega}\chi{''}(\omega)$ (red dots). For $\alpha=0.1$ (Fig.5c) the red dots fall on the blue line so the FDT is obeyed. This is consistent with the fact that for $\alpha=0.1$ there are no evaporation/nucleation events, the only noise source that violates the PDB. Once such events appear (Figs.5b and 5c) the FDT is, in both cases, weakly violated at higher frequencies and strongly violated at lower frequencies.}       \label{fig:fdt_alpha_vary}
\end{figure}

Figures \ref{fig:fdt_alpha_vary} show the fluctuation power spectra for $\alpha$ equal to $0.1$, $0.2$, and $0.3$ as blue lines, The blue shading surrounding the lines indicates the measurement error. The waiting time was kept fixed at $\tau=3.0$. 
We saw earlier that for $\alpha=0.1$ active noise plays no role so the FDT should be obeyed for this case. The red dots showing $-\frac{2}{\omega}\chi{''}(\omega)$ lie on the blue curve so the FDT is indeed obeyed. Within Landau theory, the power spectrum should be a Lorentzian with a half width equal to the relaxation rate $k(R^+)/\gamma$, which is about 0.54 (the actual half width is about 0.26) while$S(\omega)$ is indeed close to a Lorentzian. 
For $\alpha=0.2$, the role of evaporation and nucleation events is becoming significant. The low-frequency limit of $S(\omega)$ decreases by a factor of about $0.7$. Recall that Landau theory predicts that the fluctuation power actually should increase. Similarly, the relaxation rate increases while Landau theory predicts a decrease. The response term $-\chi''(\omega)/\omega$ of the FDT decreases even more than $S(\omega)$ and the FDT is clearly violated. Note that the response term $-\chi''(\omega)/\omega$ appears to have a maximum around $\omega\simeq 0.2$. For $\alpha=0.3$ (see Fig.5c), this trend increases even more. The low-frequency part of the power spectrum and the response function decreases by an additional factor of three.

\begin{figure}
    \centering
     \begin{subfigure}[b]{0.45\textwidth}
         \centering
         \includegraphics[width=\textwidth]{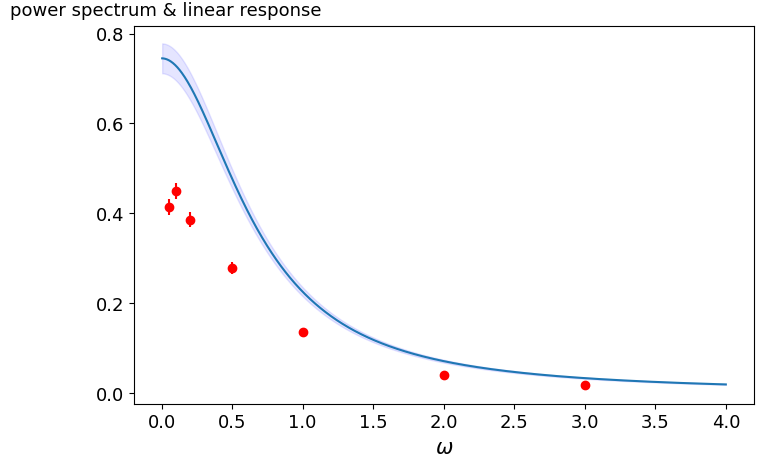}
         \caption{$\alpha = 0.3$, $\tau = 0.0$}
         \label{fig:0.3_0.0 full fdt}
     \end{subfigure}
       \begin{subfigure}[b]{0.45\textwidth}
         \centering
         \includegraphics[width=\textwidth]{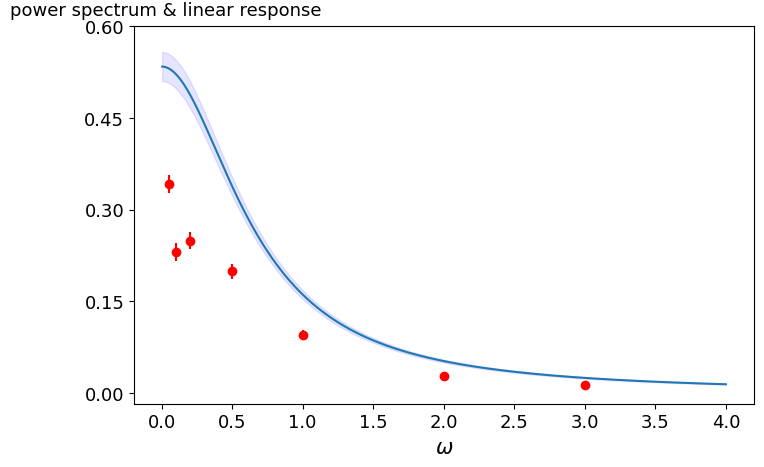}
         \caption{$\alpha = 0.3$, $\tau = 2.0$}
         \label{fig:0.3_2.0 full fdt}
     \end{subfigure}
        
    \begin{subfigure}[b]{0.45\textwidth}
    \includegraphics[width=\textwidth]
    {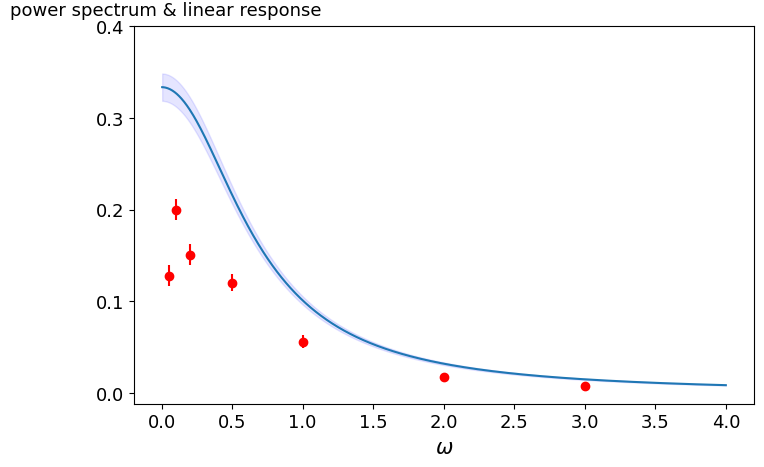}\caption{$\alpha = 0.3$, $\tau = 6.0$} \label{fig:fdt_0.3_6}
    \end{subfigure}
    \hfill
        \caption{The Fluctuation Dissipation Theorem for $\alpha$ = 0.3 and varying droplet re-nucleation waiting times ($\tau$ = 0.0, 2.0, and 6.0).}      \label{fig:fdt_tau_vary}
\end{figure}

Next, Fig.\ref{fig:fdt_tau_vary} shows the case when $\alpha$ is kept fixed at $0.3$ while the waiting time $\tau$ is varied
Increasing the waiting time for droplet regeneration reduces both $S(\omega$ and $-\frac{2}{\omega}\chi{''}(\omega)$. This is reasonable since the system spends an increasing fraction of time in a state in which it can neither fluctuate nor respond. The violation of the FDT is rather similar for the three cases. 

\section{Discussion}
 We propose a minimal model for active droplet nucleation in a confining medium characterized by two collective degrees of freedom: the droplet radius and the droplet displacement in a symmetric potential trap. Time-reversal symmetry is violated by active droplet nucleation in which new droplets are initiated only at the center of a trapping potential while droplet evaporation can take place for any displacement. In the absence of such nucleation/evaporation events, the model has a weakly first order transition characterized by an increase in the power of thermal fluctuations and an increase of the linear response susceptibility. Active droplet nucleation/evaporation suppresses the increase of the quasi-critical thermal fluctuations but it suppresses the increase of the susceptibility even more so the FDT is clearly violated at the macroscopic level. Normally, when a passive system is coupled it to an active energy source, for example when myosin motor proteins linked to an actin network are activated, then this \textit{enhances} the low-frequency power spectrum, while the linear response susceptibility is not affected \cite{Fred-Alex} while in the present case both fluctuation power and energy dissipation are weakened, though in a disproportionate fashion. It is reasonable that repeated resetting the displacement and radius to zero will reduce both low-frequency fluctuations and susceptibility but why it affects the susceptibility in a disproportionate manner is unclear. 

We focused in this paper on the secondary order parameter. Could it be that violation of the FDT by the primary order parameter would produce a more conventional scenario? Applying a thermodynamic force conjugate to the primary order parameter, say by periodic heating and cooling of a drop, would require making assumptions on how such a force would couple to the active force. Nevertheless, within the model, one can still investigate the power spectrum of radius fluctuations without having to do so. An example is shown in Fig.7.
\begin{figure}
    \centering
    \includegraphics[scale=0.4]
    {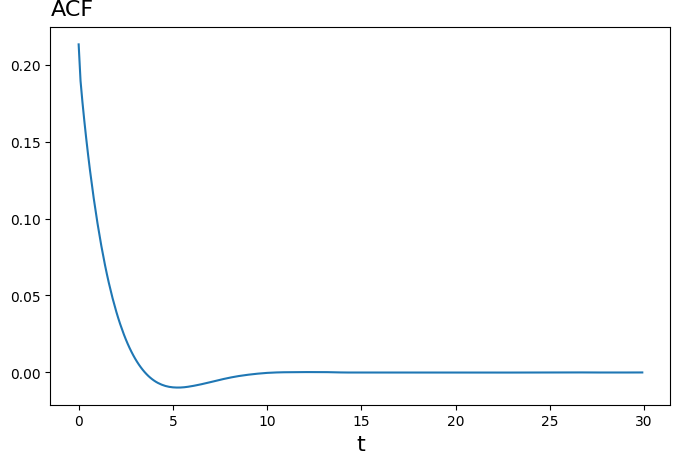}
    \caption{Auto-correlation function (ACF) of the radius $R(t)$ for $\alpha = 0.3$ and $\tau = 1.0$.} 
    \label{radi_acf_ap0.3}
\end{figure}
The radius auto-correlation function $<R(t)R(t+\Delta t)>$ shows a moderate anti-correlation for time intervals $\Delta t$ of about 3-10 time steps. This may be due to the spikes during the off-periods of Fig.3d: a non-zero value for $R(t)$ at a certain $t$ makes it very likely that $R(t)=0$ a few time steps later. The short time correlation $<R(t)R(t+1)>\simeq 0.2$ is quite small, consistent with extended intervals in the empty state with repeated but failing attempts at nucleation. This indicates that, as expected, the nucleation/evaporation events also suppress the correlation of the radius in time.

It is interesting to see what happens if one simplifies the model even further. Figure 3 shows that evaporation events are preceded by large excursions of $r(t)$ from the origin. Figure 2 shows that this will happen in particular when $r(t)$ leaves the interior of the red dashed line. This can be viewed as a \textit{first-passage} problem. Fix, for now, the radius at $R+$ so $r(t)$ is a harmonic degree of freedom. It is easy to show that the mean first passage time $\Tilde{\tau}$ is then of the order of $$(\gamma {r^*}^2/k_bT)e^{\frac{{r^*}^2}{2k(R^+) k_bT}}$$Treat the displacement degree of freedom as a random walker that is reset to the origin at random times with a mean $\Tilde{\tau}$ and have this act as a ``resetting" noise source \cite{evans2020stochastic} for the radius degree of freedom. The radius itself is treated as a harmonic variable with thermal noise plus resetting noise. This system happens to be analytically soluble \cite{gueneau2023active}. The response function of the radius is not affected while the power spectrum for radius fluctuations is enhanced so the FDT is violated. This is consistent with the intuition that active noise increases the power of the fluctuations but does not alter the response function. The displacement degree of freedom is then a \textit{resetting Brownian Walk} (rBW), where a random walker is transferred back to the origin at time-intervals chosen from some probability distribution \cite{sokolov2023linear}. Note that an rBW explicitly violates time-reversal symmetry: particles are never instantaneously translated from the origin to a point a finite distance away from the origin. It has been shown that an rBW does obey the FDT, provided the second moment of the interval time distribution function is finite \cite{sokolov2023linear}. Note that this is a second case where microscopic noise violating the PDB does not lead to failure of the FDT at the macroscopic level. The simplified system thus can not account for the observed violation of the FDT for the displacement degree of freedom. In short, it is the coupling between the two degrees of freedom that leads to the violation of the FDT in both degrees of freedom. We plan to investigate in the future violation of the FDT for the radius degree of freedom for the model but now include a thermodynamic force conjugated to the radius degree of freedom that also couples to the active source at the origin.

Finally, it must be noted that according to Figs. 1 and 2 show, that the droplet assembly free energy is in the range of just about ten $k_B T$ while the equilibrium droplet radius $R^+$ and displacement $r$ are in the range of the molecular scale $v^{1/3}$. It is only in this regime that, within the model, there is an interesting dynamical steady state with droplet nucleation and evaporation in balance. Note also that there are only a modest number of molecules per droplet so the continuum description can be questioned. It would be interesting to develop a coarse-grained Molecular Dynamics simulation model to investigate active nucleation in a confined system to explore this further.

\section{Acknowledgements}
RB would like to thank the NSF-DMR for support under CMMT Grant 1836404. CL acknowledges support from the Simons Foundation, with the Flatiron Institute operating as a division of the foundation. Special thanks are extended to Sonya Hanson and Pilar Cossio for their insightful discussions. We would also like to honor the memory of Dr.Alex Levine, whose initial guidance and inspiration were fundamental to the inception of this project. His contributions to the field continue to inspire and guide us.

\clearpage  
\bibliographystyle{unsrt}  
\bibliography{ref}  

\end{document}